\documentclass[traditabstract]{aa}
\bibpunct{(}{)}{;}{a}{}{,} 
\pdfoutput=1
\usepackage{graphicx}
\usepackage[varg]{txfonts}
\usepackage{color}
\usepackage{lscape}
\usepackage{pdfpages}

\newcommand{\srm}{\scriptscriptstyle\rm}

\newcommand{\bgal}{b_{\srm II}}

\begin{document}

\title{Optical linear polarization measurements of quasars obtained with the 3.6m telescope at the La Silla Observatory.\thanks{Based on observations made with the ESO 3.6m Telescope at the La Silla Observatory under program ID 071.B-0460, 079.A-0625, 080.A-0017.}}
\author{D. Hutsem\'ekers\inst{1,}\thanks{Senior Research Associate F.R.S.-FNRS.},
        P. Hall\inst{2},
        D. Sluse\inst{1}
        }
\institute{
    Institut d'Astrophysique et de G\'eophysique,
    Universit\'e de Li\`ege, All\'ee du 6 Ao\^ut 19c, B5c,
    4000 Li\`ege, Belgium
    \and 
    Department of Physics and Astronomy, York University, Toronto, Ontario M3J 1P3, Canada
    }
\date{Received ; accepted: }
\titlerunning{Quasar polarization measurements} 
\authorrunning{D. Hutsem\'ekers et al.}
\abstract{We report 192 previously unpublished optical linear polarization measurements of quasars obtained in April 2003, April 2007, and October 2007 with the European Southern Observatory Faint Object Spectrograph and Camera (EFOSC2) instrument attached to the 3.6m telescope at the La Silla Observatory. Each quasar was observed once. Among the 192 quasars, 89 have a polarization degree $p \geq 0.6\%$, 18 have $p \geq 2\%$, and two have $p \geq 10\%$.
}
\maketitle
%
%
%
\section{Introduction}
\label{sec:intro}

The linear polarization of optical light is an important feature in the study of quasars and other active galactic nuclei (AGN). Usually attributed to scattering, polarization is directly related to the object symmetry axis, and is at the heart of AGN unification models. 

In the present paper, we report new optical linear polarization measurements of quasars obtained with EFOSC2, the European Southern Observatory (ESO) Faint Object Spectrograph and Camera instrument attached to the 3.6m telescope at the La Silla Observatory. Although the observations were designed for various scientific goals, the quality of the data is homogeneous. In Sect.~\ref{sec:obs} we describe the observing procedure. Data reduction and measurements are summarized in Sect.~\ref{sec:reduc}. The online table with the final measurements is outlined in Sect.~\ref{sec:data}.

\section{Observations}
\label{sec:obs}

The polarimetric observations were carried out in April 2003, April 2007, and October 2007 at the European Southern Observatory, La Silla, using the 3.6m telescope equipped with EFOSC2 attached to the Cassegrain focus. Linear polarimetry is performed by inserting a Wollaston prism into the parallel beam, which splits the incoming light rays into two orthogonally polarized beams. Each object in the field has therefore two orthogonally polarized images on the charge-coupled device (CCD) detector, separated by 20$\arcsec$. To avoid image overlapping, one puts at the telescope focal plane a special mask made of alternating transparent and opaque parallel strips whose widths correspond to the splitting.  The final CCD image then consists of alternate orthogonally polarized strips of the sky, two of them containing the polarized images of the object itself \citep{1989diSerego,1997diSerego,1999Lamy}.  Since the two orthogonally polarized images of the object are simultaneously recorded, the polarization measurements do not depend on variable atmospheric transparency, or seeing. In order to derive the normalized Stokes parameters $q$ and $u$, four frames are obtained with the half-wave plate (HWP) at four different position angles (0$\degr$, 22.5$\degr$, 45$\degr$, and 67.5$\degr$). While only two different orientations of the HWP are sufficient to measure the linear polarization, the two additional orientations allow us to remove most of the instrumental polarization \citep{1989diSerego}. Polarized and unpolarized standard stars were observed to unambiguously fix the zero-point of the polarization position angle, to estimate the instrumental polarization, and to check the whole observing and reduction process.

Most targets are quasars with redshifts between one and three, and V magnitudes between 17 and 19. They were mostly selected from amongst broad absorption line (BAL), radio-loud, or red quasars that are more likely to be significantly polarized. All observations but one were obtained through the Bessel V filter (ESO\# 641), the Bessel R filter (ESO\# 642), and the Gunn i filter (ESO\# 705), with typical exposure times per frame ranging between one and ten minutes. One faint target was observed unfiltered (in ``white light''). The CCD\#40 mounted on EFOSC2 is a 2k$\times$2k CCD with a pixel size of 15 $\mu$m corresponding to 0.158\arcsec\ on the sky in the 1$\times$1 binning mode. 

\section{Data reduction and measurements}
\label{sec:reduc}

\begin{table*}[t]
\caption[ ]{Observed standard stars.}
\label{tab:std}
\centering   
\begin{tabular}{lll}
\hline\hline
Date & Polarized & Unpolarized\\ 
yyyy-mm-dd & &\\
\hline\\
2003-04-04  & HD155197, HD298383 & HD94851 \\
2003-04-05  & HD155197           & HD64299 \\
2003-04-06  & HD298383           & HD154892 \\
2003-04-07  & HD126593           & HD64299 \\
2003-04-08  & HD155197           &  \\
2007-04-22  & HD155197, Ve6$-$23 &   \\
2007-04-23  & HD155197 & WD1615$-$154   \\
2007-10-04  & HD155197  &    \\
2007-10-05  & BD$-$14$\degr$4922, NGC2024$-$1, HD316232     &  HD64299, WD1620$-$391   \\
2007-10-06  & HD155197   &  WD0310$-$688  \\
2007-10-07  & BD$-$14$\degr$4922, NGC2024$-$1, BD$-$12$\degr$5133  & HD64299    \\
2007-10-08  & HD316232, Ve6$-$23  &  WD0310$-$688, WD1620$-$391   \\
\hline
\end{tabular}
\tablebib{\citet{1990Turnshek,2007Fossati}.} 
\end{table*} 

\begin{table}[t]
\caption[ ]{Residual polarization from field stars.}
\label{tab:stars}
\centering   
\begin{tabular}{lrrrr}
\hline\hline
Observing run & $\overline{q}_{\star}$ \ & $\overline{u}_{\star}$ \ &  $\overline{\sigma}_{\star}$ \ & $n_{\star}$\\
   yyyy-mm    &     (\%) & (\%) & (\%) &\\
\hline \\
2003-04 & $-$0.10 & $+$0.10 & 0.14 & 64 \\
2007-04 & $-$0.11 & $+$0.14 & 0.14 & 31 \\
2007-10 & $+$0.03 & $-$0.12 & 0.14 & 86 \\
\hline
\end{tabular}
\end{table} 

The $q$ and $u$ Stokes parameters are computed from the measurement of the integrated intensity ratios between the upper and lower orthogonally polarized images of the object, for the four different orientations of the half-wave plate. They are calculated with respect to the instrumental reference frame according to
\begin{eqnarray}
q & = & \frac{R_q - 1}{R_q + 1}  \hspace{0.5cm} \mbox{where} \hspace{0.5cm} 
R_q^2  = \frac{I_{\srm 0}^{\srm u}/I_{\srm 0}^{\srm l}}
       {I_{\srm 45}^{\srm u}/I_{\srm 45}^{\srm l}}, \;\;\; \rm{and}  \nonumber\\
 & &\\
u & = & \frac{R_u - 1}{R_u + 1}  \hspace{0.5cm} \mbox{where} \hspace{0.5cm} 
R_u^2  = \frac{I_{\srm 22.5}^{\srm u}/I_{\srm 22.5}^{\srm l}}
       {I_{\srm 67.5}^{\srm u}/I_{\srm 67.5}^{\srm l}},\nonumber 
\end{eqnarray}
where $I^{\srm u}$ and $I^{\srm l}$ refer to the intensities (electron counts) integrated over the upper and lower images of the object, respectively. The photometric measurements were done using the procedures described in \citet{1999Lamy} and \citet{2005Sluse}. The positions of the  upper and lower images are measured at subpixel precision by fitting two-dimensional Gaussian profiles. The intensities are then integrated in circles centered on the upper and lower images, and the Stokes parameters are computed for various values of the aperture radius. Since the Stokes parameters are found to be stable against radius changes, we adopt a fixed aperture radius of $3.0 \times [(2 \ln 2)^{-1/2}\, \rm{HWHM}],$ where HWHM represents the mean half-width at half-maximum of the two-dimensional Gaussian profile. In a few cases, the Stokes parameters strongly fluctuate when changing the aperture radius, making their measurement unreliable. The combination of measurements using four frames obtained with different HWP orientations removes most of the instrumental polarization, and corrects the effects of image distortions introduced by the HWP \citep{1989diSerego,1999Lamy}. The uncertainties $\sigma_q$ and $\sigma_u$ are evaluated by computing the errors on the intensities $I^{\srm u}$ and $I^{\srm l}$ from the read-out noise and the photon noise in the object and the sky background, and then by propagating these errors. Typical uncertainties are around 0.2\% for either $q$ or $u$.

A zero-point angle offset correction, filter dependent, is then applied to the quasar normalized Stokes parameters $q$ and $u$ in order to convert the polarization angle measured in the instrumental reference frame to the equatorial reference direction. This angle offset is determined using polarized standard stars observed every night in each filter (Table~{\ref{tab:std}}). For all stars observed during a given run and within a given filter, the values of the offset agree within 1\degr .
The polarization of unpolarized standard stars (Table~{\ref{tab:std}}) is typically around 0.10 $\pm$ 0.05 \% for all runs indicating that the instrumental polarization is small and essentially removed by the observing procedure.

\begin{table*}[t]
\caption[ ]{Polarization of distant stars.}
\label{tab:halo}
\centering   
\begin{tabular}{lccccccccccr}
\hline\hline
Reference-number & RA & DEC &  Distance &  Obs. Date & $q$ & $u$ & $p$  & $\sigma_p$ & $p_0$ & $\theta$ & $\sigma_{\theta}$  \\
     & h m s    &  $\degr$ $\arcmin$ $\arcsec$ & kpc & & \% & \% & \% & \% & \degr & \degr & \degr \\
\hline \\
Beers-891      & 12 47 17.77 & $-$30 41 33.8 & 10.7 &    2007-04-23 &  $-$0.03 &  $-$0.16  &   0.16  &  0.14  &  0.10  &  129 &  40\\
Beers-923      & 13 01 06.56 & $+$27 35 06.1 & 15.9 &    2007-04-23 &  $+$0.16 &  $-$0.11  &   0.19  &  0.15  &  0.14  &  163 &  31\\
Beers-1035     & 13 17 12.48 & $-$05 04 52.2 & 10.8 &    2007-04-22 &  $+$0.01 &  $+$0.18  &   0.18  &  0.14  &  0.14  &  ~44 &  30\\
Beers-1670     & 21 31 54.40 & $-$42 35 41.0 & 19.8 &    2007-10-06 &  $+$0.12 &  $-$0.19  &   0.22  &  0.15  &  0.18  &  152 &  24\\
Clewley-CMTF010& 22 10 32.80 & $-$16 12 09.8 & 22.4 &    2007-10-08 &  $-$0.23 &  $+$0.04  &   0.23  &  0.20  &  0.15  &  ~84 &  39\\
\hline
\end{tabular}
\tablebib{\citet{2000Beers,2004Clewley}.} 
\end{table*} 

Since on most frames field stars are simultaneously recorded, one can use them to estimate the residual instrumental polarization and/or interstellar polarization. While a frame-by-frame correction of the quasar Stokes parameters is in principle possible, it is hazardous since we are never sure that the polarization of field stars correctly represents the interstellar polarization that could affect distant quasars. Moreover, the field stars can be fainter than the quasar so that a frame-by-frame correction would introduce uncertainties on the quasar polarization larger than the residual polarization itself. We then compute the weighted average ($\overline{q}_{\star}$ and $\overline{u}_{\star}$) and dispersion ($\overline{\sigma}_{\star}$) of the normalized Stokes parameters of field stars, considering the $n_{\star}$ frames with suitable field stars obtained during a given run. These values  are given in Table~{\ref{tab:stars}}. Frames centered on quasars and distant stars (Table~\ref{tab:halo}) are considered. We usually measure a single field star per frame; in some cases this star is made up of the combination of two to three fainter stars observed on the same frame. Only stellar polarizations with uncertainties  $\sigma_q$ and $\sigma_u$ better than 0.3\% are used in the average. The small values and dispersions of the average Stokes parameters reported in Table~{\ref{tab:stars}} confirm the low level of uncorrected instrumental polarization.

In order to minimize systematic errors in the sample, we conservatively take this residual instrumental and/or averaged interstellar polarization into account by subtracting the systematic $\overline{q}_{\star}$ and $\overline{u}_{\star}$ from the measured $q$ and $u$, and by adding quadratically $\overline{\sigma}_{\star}$ to their errors. Then, from the corrected $q$ and $u$ values, the polarization degree is evaluated using $p = (q^2+u^2)^{1/2}$ and the associated error using $\sigma_p  \simeq \sigma_q \simeq \sigma_u$.  In addition, $p$ must be corrected for the statistical bias inherent to the fact that $p$ is always a positive quantity.  The debiased value $p_{0}$ of the polarization degree is obtained by using the \citet{1974Wardle} estimator, which was found to be a reasonably good estimator of the true polarization degree \citep{1985Simmons}. The polarization position angle $\theta$ is obtained by solving the equations $q = p\cos 2\theta$ and $u = p \sin 2\theta$. The uncertainty of the polarization position angle $\theta$ is estimated from the standard \citet{1962Serkowski} formula, where the debiased value $p_{0}$ is conservatively used instead of $p$, that is, $\sigma_{\theta} = 28\fdg65\, \sigma_p / p_{0}$ \citep[see also][]{1974Wardle}. Due to the HWP chromatism over broad-band filters, an additional error $\leq 2-3\degr$ on $\theta$ should be accounted for (cf. the wavelength dependence of the polarization position angle offset in \citealt{1997diSerego}).

These procedures, the values of the residual polarization, and the distribution of field star polarization are similar to those reported in \citet{2005Sluse}. We refer to that paper for more details and an exhaustive discussion of the effect of the various corrections. As a conclusion, the polarization of field stars is most often $\leq 0.3\%$ so that virtually every quasar with a polarization degree higher than $0.6\%$ is intrinsically polarized, in agreement with previous studies \citep{1990Berriman,2000Lamy}.

Since the distance of field stars in unknown, we have measured the V-band polarization of a few very distant stars ($d > $ 10 kpc) to further check the amount of interstellar polarization in the direction of our targets. These measurements are reported in Table~\ref{tab:halo}. All these stars appear to have low polarization. Although the sample is small, this confirms that, on average, contamination by interstellar polarization should be unimportant for quasars at high galactic latitudes ($|\bgal| > 30\degr$) and with polarization degrees higher than 0.6\%.

\section{Polarization data}
\label{sec:data}

The full Table~\ref{tab:qsos}, available at the Strasbourg astronomical data center (CDS), summarizes the polarization measurements obtained for 192 different quasars (72 in April 2003, 28 in April 2007, and 92 in October 2007). Unreliable measurements were discarded. Eighty-nine quasars have $p \geq 0.6\%$, 18 have $p \geq 2\%$, and two have $p \geq 10\%$. Column~(1) gives the quasar name from the  NASA/IPAC extragalactic database (NED), Columns~(2) and~(3) the equatorial coordinates (J2000), Column~(4) the redshift $z$, Column~(5) the filter used (V, R, i, and W for no filter), and Column~(6) the date of observation (year-month-day). Columns~(7) and~(8) give the normalized Stokes parameters $q$ and $u$ in percent corrected for the systematic residual polarization given in Table~\ref{tab:stars}. The normalized Stokes parameters are given in the equatorial reference frame.  Columns~(9) and~(10) give the polarization degree $p$ and its error $\sigma_p$ in percent. Column~(11) gives the debiased polarization degree $p_0$ in percent. Columns~(12) and~(13) give the polarization position angle $\theta$ east-of-north and its error $\sigma_{\theta}$, in degrees. When $p < \sigma_p$, the polarization angle is undefined and its value put to 999. Finally ``C?'' in column~(14) indicate quasars with $p > 0.6\%$ and for which the polarization of the field stars is comparable or higher, that is objects whose polarization has possibly been contaminated by interstellar polarization.

Among the eight objects with $p \geq 3\%$ reported in Table~\ref{tab:qsos}, three of them were already known for their high linear polarization: $[$HB89$]$~0219-164 \citep{1990Mead}, WISE~J125908.45-231038.6 \citep{2005Sluse}, and $[$HB89$]$~2155-152 \citep{1984Wardle}. Moreover, the linear polarization of WISE~J125908.45-231038.6 and $[$HB89$]$~2155-152, uncorrected for the systematic polarization given in Table~\ref{tab:stars}, has been discussed in \citet{2010Hutsemekers} together with circular polarization measurements. Among the five remaining objects with first-time polarization measurements, two are BAL quasars: SDSS~J145603.07+011445.4 \citep{2002Hall} and SDSS~J232550.73-002200.3 \citep{2006Trump}. $[$HB89$]$~1243-072, WISE~J131250.90-042449.9 and PKS~2140-43 are Parkes radio sources, the first two also belonging to the Fermi Gamma-ray Space Telescope source catalog \citep{2012Nolan}.

\begin{sidewaystable}
\caption[ ]{Polarization of quasars.}
\label{tab:qsos}
\centering   
\begin{tabular}{lcccccrrrrrrrc}
\hline\hline
Name & RA    & DEC                         & $z$ & Filter & Obs. Date & $q$ & $u$ & $p$  & $\sigma_p$ & $p_0$  & $\theta$ & $\sigma_{\theta}$  & \\
     & h m s & $\degr$ $\arcmin$ $\arcsec$ &     &        &           & \%  & \%  & \%   & \%         & \degr  & \degr    & \degr            & \\
\hline \\
$[$HB89$]$ 0219-164              &  02 22 00.72 & $-$16 15 16.5 &  0.700000 &  V &  2007-10-09 & $-$3.07 & $-$3.91 &   4.97 &   0.53 &   4.94 &   116 &    3 &   \\  
$[$HB89$]$ 1243-072              &  12 46 04.23 & $-$07 30 46.6 &  1.286000 &  V &  2007-04-23 &    8.35 & $-$3.45 &   9.03 &   0.62 &   9.01 &   169 &    2 &   \\     
WISE J125908.45-231038.6         &  12 59 08.46 & $-$23 10 38.7 &  0.481000 &  V &  2007-04-22 &   12.74 & $-$8.99 &  15.59 &   0.21 &  15.59 &   162 &    1 &   \\     
WISE J131250.90-042449.9         &  13 12 50.90 & $-$04 24 49.9 &  0.824900 &  V &  2007-04-23 &    0.13 &    8.70 &   8.70 &   0.51 &   8.69 &    45 &    2 &  \\      
PKS 1420-27                      &  14 22 49.23 & $-$27 27 55.9 &  0.985000 &  V &  2007-04-22 & $-$0.14 &    0.04 &   0.15 &   0.19 &   0.00 &   999 &  999 &  \\      
$[$HB89$]$ 1424-118              &  14 27 38.10 & $-$12 03 50.0 &  0.802945 &  V &  2007-04-23 & $-$0.96 & $-$0.34 &   1.02 &   0.18 &   1.00 &   100 &    5 & C?  \\
SDSS J145603.07+011445.4         &  14 56 03.08 & $+$01 14 45.5 &  2.363504 &  W &  2003-04-08 & $-$7.74 & $-$5.23 &   9.34 &   0.55 &   9.32 &   107 &    2 &  \\   
PKS 2140-43                      &  21 43 33.39 & $-$43 12 47.9 &  0.650000 &  V &  2007-10-09 & $-$5.19 &    2.52 &   5.77 &   0.17 &   5.77 &    77 &    1 &  \\   
$[$HB89$]$ 2155-152              &  21 58 06.28 & $-$15 01 09.3 &  0.672000 &  V &  2007-04-22 & $-$3.56 &   17.14 &  17.51 &   0.51 &  17.50 &    51 &    1 &  \\      
SDSS J232550.73-002200.3         &  23 25 50.73 & $-$00 22 00.4 &  1.010857 &  V &  2007-10-08 &    1.96 & $-$4.00 &   4.45 &   0.22 &   4.44 &   148 &    1 & \\
\hline
\end{tabular}
\tablefoot{This table shows the polarization measurements for 10 quasars. It includes the eight objects  with $p \geq 3\%$. The complete table is available electronically at CDS. }
\end{sidewaystable}

\begin{acknowledgements}
This research has made use of the NASA/IPAC Extragalactic Database (NED), which is operated by the Jet Propulsion Laboratory, California Institute of Technology, under contract with the National Aeronautics and Space Administration. 
\end{acknowledgements}

\bibliographystyle{aa}
\bibliography{references}

\begin{thebibliography}{19}
\expandafter\ifx\csname natexlab\endcsname\relax\def\natexlab#1{#1}\fi

\bibitem[{{Beers} {et~al.}(2000){Beers}, {Chiba}, {Yoshii}, {Platais},
  {Hanson}, {Fuchs}, \& {Rossi}}]{2000Beers}
{Beers}, T.~C., {Chiba}, M., {Yoshii}, Y., {et~al.} 2000, \aj, 119, 2866

\bibitem[{{Berriman} {et~al.}(1990){Berriman}, {Schmidt}, {West}, \&
  {Stockman}}]{1990Berriman}
{Berriman}, G., {Schmidt}, G.~D., {West}, S.~C., \& {Stockman}, H.~S. 1990,
  \apjs, 74, 869

\bibitem[{{Clewley} {et~al.}(2004){Clewley}, {Warren}, {Hewett}, {Norris}, \&
  {Evans}}]{2004Clewley}
{Clewley}, L., {Warren}, S.~J., {Hewett}, P.~C., {Norris}, J.~E., \& {Evans},
  N.~W. 2004, \mnras, 352, 285

\bibitem[{{di Serego Alighieri}(1989)}]{1989diSerego}
{di Serego Alighieri}, S. 1989, in European Southern Observatory Conference and
  Workshop Proceedings, Vol.~31, ESO/ST-ECF Data Analysis Workshop, ed. P.~J.
  {Grosb{\o}l}, F.~{Murtagh}, \& R.~H. {Warmels}, 157--160

\bibitem[{{di Serego Alighieri}(1997)}]{1997diSerego}
{di Serego Alighieri}, S. 1997, {Polarimetry with large telescopes.}, ed. J.~M.
  {Rodr{\'{\i}}guez Espinosa}, A.~{Herrero}, \& F.~{S{\'a}nchez}, 287--329

\bibitem[{{Fossati} {et~al.}(2007){Fossati}, {Bagnulo}, {Mason}, \& {Landi
  Degl'Innocenti}}]{2007Fossati}
{Fossati}, L., {Bagnulo}, S., {Mason}, E., \& {Landi Degl'Innocenti}, E. 2007,
  in Astronomical Society of the Pacific Conference Series, Vol. 364, The
  Future of Photometric, Spectrophotometric and Polarimetric Standardization,
  ed. C.~{Sterken}, 503

\bibitem[{{Hall} {et~al.}(2002){Hall}, {Anderson}, {Strauss}, {York},
  {Richards}, {Fan}, {Knapp}, {Schneider}, {Vanden Berk}, {Geballe}, {Bauer},
  {Becker}, {Davis}, {Rix}, {Nichol}, {Bahcall}, {Brinkmann}, {Brunner},
  {Connolly}, {Csabai}, {Doi}, {Fukugita}, {Gunn}, {Haiman}, {Harvanek},
  {Heckman}, {Hennessy}, {Inada}, {Ivezi{\'c}}, {Johnston}, {Kleinman},
  {Krolik}, {Krzesinski}, {Kunszt}, {Lamb}, {Long}, {Lupton}, {Miknaitis},
  {Munn}, {Narayanan}, {Neilsen}, {Newman}, {Nitta}, {Okamura}, {Pentericci},
  {Pier}, {Schlegel}, {Snedden}, {Szalay}, {Thakar}, {Tsvetanov}, {White}, \&
  {Zheng}}]{2002Hall}
{Hall}, P.~B., {Anderson}, S.~F., {Strauss}, M.~A., {et~al.} 2002, \apjs, 141,
  267

\bibitem[{{Hutsem{\'e}kers} {et~al.}(2010){Hutsem{\'e}kers}, {Borguet},
  {Sluse}, {Cabanac}, \& {Lamy}}]{2010Hutsemekers}
{Hutsem{\'e}kers}, D., {Borguet}, B., {Sluse}, D., {Cabanac}, R., \& {Lamy}, H.
  2010, \aap, 520, L7

\bibitem[{{Lamy} \& {Hutsem{\'e}kers}(1999)}]{1999Lamy}
{Lamy}, H. \& {Hutsem{\'e}kers}, D. 1999, The Messenger, 96, 25

\bibitem[{{Lamy} \& {Hutsem{\'e}kers}(2000)}]{2000Lamy}
{Lamy}, H. \& {Hutsem{\'e}kers}, D. 2000, \aaps, 142, 451

\bibitem[{{Mead} {et~al.}(1990){Mead}, {Ballard}, {Brand}, {Hough}, {Brindle},
  \& {Bailey}}]{1990Mead}
{Mead}, A.~R.~G., {Ballard}, K.~R., {Brand}, P.~W.~J.~L., {et~al.} 1990, \aaps,
  83, 183

\bibitem[{{Nolan} {et~al.}(2012){Nolan}, {Abdo}, {Ackermann}, {Ajello},
  {Allafort}, {Antolini}, {Atwood}, {Axelsson}, {Baldini}, {Ballet}, \&
  et~al.}]{2012Nolan}
{Nolan}, P.~L., {Abdo}, A.~A., {Ackermann}, M., {et~al.} 2012, \apjs, 199, 31

\bibitem[{{Serkowski}(1962)}]{1962Serkowski}
{Serkowski}, K. 1962, Advances in Astronomy and Astrophysics, 1, 289

\bibitem[{{Simmons} \& {Stewart}(1985)}]{1985Simmons}
{Simmons}, J.~F.~L. \& {Stewart}, B.~G. 1985, \aap, 142, 100

\bibitem[{{Sluse} {et~al.}(2005){Sluse}, {Hutsem{\'e}kers}, {Lamy}, {Cabanac},
  \& {Quintana}}]{2005Sluse}
{Sluse}, D., {Hutsem{\'e}kers}, D., {Lamy}, H., {Cabanac}, R., \& {Quintana},
  H. 2005, \aap, 433, 757

\bibitem[{{Trump} {et~al.}(2006){Trump}, {Hall}, {Reichard}, {Richards},
  {Schneider}, {Vanden Berk}, {Knapp}, {Anderson}, {Fan}, {Brinkman},
  {Kleinman}, \& {Nitta}}]{2006Trump}
{Trump}, J.~R., {Hall}, P.~B., {Reichard}, T.~A., {et~al.} 2006, \apjs, 165, 1

\bibitem[{{Turnshek} {et~al.}(1990){Turnshek}, {Bohlin}, {Williamson}, {Lupie},
  {Koornneef}, \& {Morgan}}]{1990Turnshek}
{Turnshek}, D.~A., {Bohlin}, R.~C., {Williamson}, II, R.~L., {et~al.} 1990,
  \aj, 99, 1243

\bibitem[{{Wardle} \& {Kronberg}(1974)}]{1974Wardle}
{Wardle}, J.~F.~C. \& {Kronberg}, P.~P. 1974, \apj, 194, 249

\bibitem[{{Wardle} {et~al.}(1984){Wardle}, {Moore}, \& {Angel}}]{1984Wardle}
{Wardle}, J.~F.~C., {Moore}, R.~L., \& {Angel}, J.~R.~P. 1984, \apj, 279, 93

\end{thebibliography}

\clearpage
\includepdf[pages=1]{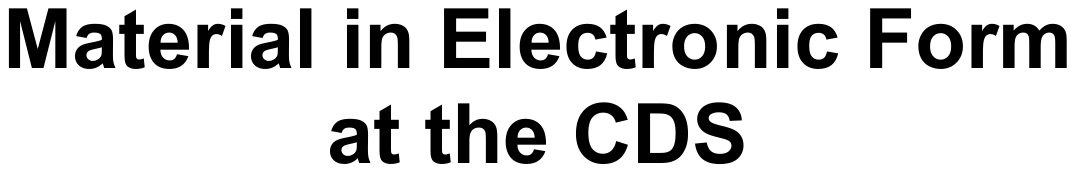}
Table 4 : Polarization of 192 quasars
\includepdf[pages=1]{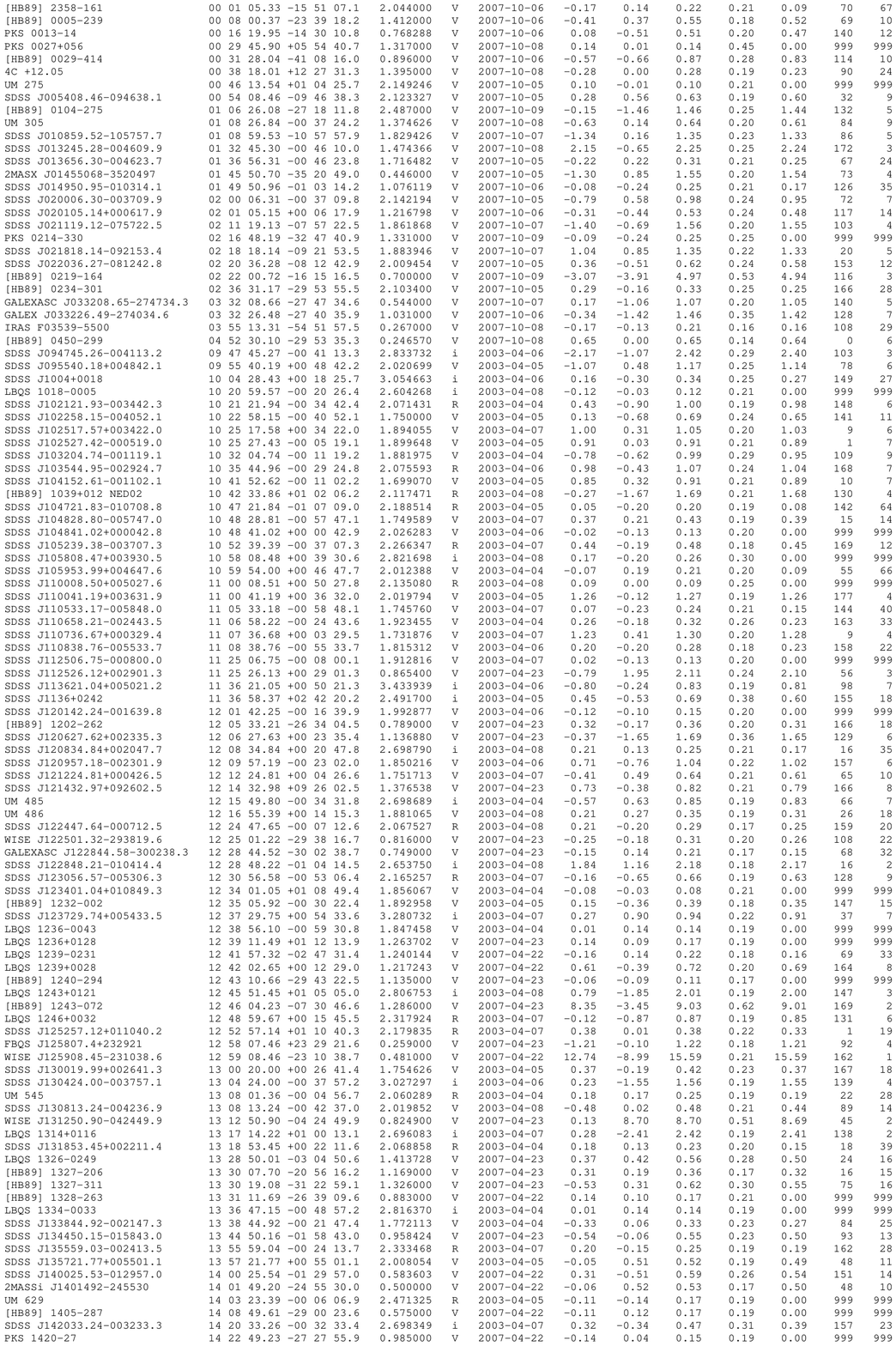}
\includepdf[pages=2]{table4.pdf}

\end{document}